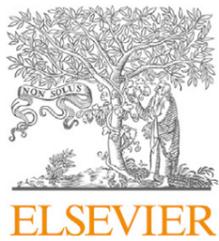
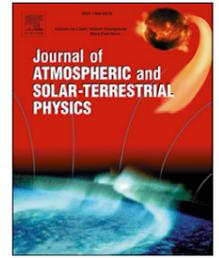
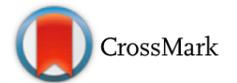

# Planetary wave-like oscillations in the ionosphere retrieved with a longitudinal chain of ionosondes at high northern latitudes

Nora H. Stray [a,b,c,*], Patrick J. Espy [a,b]

[a] Department of Physics, NTNU, 7491 Trondheim, Norway
[b] Birkeland Centre for Space Science, 5020 Bergen, Norway
[c] Teknova AS, Tordenskjolds Gate 9, 4612 Kristiansand, Norway

ABSTRACT

This paper examines the influence of neutral dynamics on the high latitude ionosphere. Using a longitudinal chain of ionosondes at high northern latitudes (52°–65° N), planetary wave-like structures were observed in the spatial structure of the peak electron density in the ionosphere. Longitudinal wavenumbers $S_0$, $S_1$ and $S_2$ have been extracted from these variations of the F layer. The observed wave activity in wavenumber one and two does not show any significant correlation with indices of magnetic activity, suggesting that this is not the primary driver. In addition, the motion of the $S_1$ ionospheric wave structures parallels that of the $S_1$ planetary waves observed in the winds of the mesosphere-lower-thermosphere derived from a longitudinal array of SuperDARN meteor-radar wind measurements. The time delay between the motions of the wave structures would indicate a indirect coupling, commensurate with the diffusion to the ionosphere of mesospheric atomic oxygen perturbations.

## 1. Introduction

There has long been an interest in the sources of ionospheric variability, particularly how this variability may be driven by the dynamics of the neutral atmosphere. Many investigations have found planetary wave-like oscillation in ionospheric parameters, particularly $f_0F2$, at periods between 2 and 20 days (e.g. Altadill, 1996, 2000; Borries et al., 2007; Forbes and Leveroni, 1992; Pancheva and Lysenko, 1988; Pancheva and Mukhtarov, 2012; Polekh et al., 2011; Shalimov et al., 2006), leading to the suggestion that planetary waves in the lower atmosphere are coupling to the ionospheric electron density. The presence of these oscillations is puzzling since model investigations show that planetary waves with these periods do not directly propagate above approximately 110 km (Forbes, 1995; Hagan et al., 1993; Pogoreltsev et al., 2007). Several potential mechanisms have been proposed to explain how planetary waves in the lower atmosphere could force these planetary wave-like oscillations in the ionosphere. These include planetary wave modulation of upward propagating tides or gravity waves imprinting that variability in the F-region, modulation of the $O/N_2$ density near the turbopause changing the F-region recombination rates, modulation of the E-region winds and dynamo-induced electric fields that modify vertical plasma drifts, or forcing from above by changes in solar EUV or solar wind pressure creating global scale patterns (Borries and Hoffmann, 2010, and references therin).

Many measurements of planetary wave-like oscillations in the ionosphere are based on temporal variations of specific wave modes with characteristic periods (e.g. 2-day or 16-day oscillations) in the ionosphere and the mesosphere or stratosphere. These rely on a specific wave mode being dominant at one station, or a series of stations, for several wave periods (e.g. Altadill, 2000; Forbes and Zhang, 1997; Pancheva et al., 1994). In the latter case, the phase progression between stations is used to infer the longitudinal wave component (e.g. wavenumber one, or $S_1$, etc.). The difficulty with this approach is that the velocity of these waves can be Doppler shifted by the background wind (Forbes, 1995; Laštovička et al., 2003), leading to unstable temporal periods and the possibility of the period-wavenumber relationship being incorrectly interpreted as a particular wave mode. It also excludes the identification of any stationary waves (Kleinknecht et al., 2014). In addition, in the presence of strong vertical wind gradients, individual temporal frequencies can be Doppler shifted to different frequencies at different levels, and some could be blocked by the Charney-Drazen criteria from ever reaching higher levels depending upon their horizontal propagation speed (Charney and Drazin, 1961). Even when correctly identified, the short periods of time for which a particular wave dominates the wave field at all levels preclude an examination of the seasonal variation (Laštovička et al., 2003).

* Corresponding author. Teknova AS, Tordenskjolds Gate 9, 4612 Kristiansand, Norway.
    E-mail address: nora.stray@teknova.no (N.H. Stray).

Later measurements of planetary wave-like oscillations in the ionosphere minimize this ambiguity by utilizing multi-satellite measurements to fit both the temporal period and spatial wavenumber for the individual component waves (e.g. Borries et al., 2007; Pancheva and Mukhtarov, 2012). However, these studies may rely on single satellite observations for the planetary waves in the lower atmosphere where spatial-temporal aliasing can lead to misinterpretation (Kleinknecht et al., 2014). In either case, each individual temporal mode of a given spatial wavenumber component of the ionospheric and lower atmospheric oscillation is compared separately in order to examine the possible coupling between the two regions. The difficulty with this procedure is that many of the mechanisms proposed for communicating the planetary wave information from the lower atmosphere into the ionosphere depend on the net wind or density structure formed by the superposition of all the temporal modes. That is, all the different temporal modes with different wave velocities, and hence temporal periods, of a wavenumber one component will superpose and form a total wavenumber one wave (e.g. Palo et al., 2005). The same is true for higher spatial wavenumbers. The amplitude and phase of this spatial component will change as the different velocity modes change, but it will maintain its wave one structure. Thus, it is this net wave-one structure that forms the background wind and density structure that interacts with tidal or gravity waves, and it is this structure that would create longitudinal structure in the upper atmospheric $O/N_2$ ratios or plasma drifts.

Here, we extract the total longitudinal wavenumber structures in the ionosphere as well as in the upper mesosphere-lower thermosphere (MLT) in order to examine whether the net phase velocities of the waves in the two regions are related. To extract the planetary wave-like structures in the ionosphere, the $f_0F2$ parameter from a longitudinal chain of ionosondes covering a limited, 13° latitude range is used. Data from a chain of ionosondes, in contrast to a single station, can be used to observe the spatial structure of planetary wave-like oscillations in the F-region and how it evolves in time. The spatial structure, representing the superposition of all temporal modes, will be stable with regard to Doppler shifting and the limited persistence of individual temporal modes (Laštovička et al., 2003). Thus, spatial mode analysis is used extensively to trace the vertical propagation of planetary waves in the strong wind gradients of the troposphere and stratosphere (e.g. Plumb, 2010, and references therein). Since observations from all the stations are made at the same time, there is no problem with aliasing spatial and temporal information that can occur with non-coincident satellite measurements. The technique used to extract the spatial wave information is similar to that described in Kleinknecht et al. (2014) for the meteor winds from a longitudinal chain of Super Dual Auroral Radar Network (SuperDARN) radars (Greenwald et al., 1985, 1995). This same technique is used to extract planetary-wave amplitudes and phases for spatial structures with wavenumbers one and two from the SuperDARN neutral meridional winds at 95 km for comparison with the ionospheric waves. The resulting structures and the temporal evolution of their phase velocities can be compared to ascertain whether the MLT waves are related to the ionospheric oscillations.

## 2. Data

The parameter $f_0F2$ (MHz) from the ionosonde is proportional to the peak electron density, $N_mF2$ ($m^{-3}$), at the F2 region peak near 250 km using the standard formula $N_mF2 = 1.24 \cdot 10^{10} \cdot (f_0F2)$. Hence, electron density variations associated with planetary wave dynamics will modulate $f_0F2$ (Altadill, 1996, 2000; Forbes and Leveroni, 1992; Pancheva and Lysenko, 1988). In order to characterize the longitudinal variation, and hence the zonal wavenumber of the disturbances, a chain of ionosondes within a latitude band between 52° and 65° N has been used. Given the latitudinal and temporal coverage of the stations, it is possible to extract planetary wave-like disturbances with longitudinal wavenumber 0, 1 and 2 in the F-region during 2001. The $f_0F2$ values were taken from the Space Physics Interactive Data Resource (SPIDR), http://spidr.ionosonde.net/spidr/ionoInventory.do. The list of the ionosondes included in this study, shown in Table 1, consists of all ionosonde stations located between 52° and 65° N with suffcient $f_0F2$ records for 2001 that were available on SPIDR.

The sampling of the $f_0F2$ values may vary anywhere between 15 min and 1 h from station to station. In order to maintain uniformity in the fitting, all observations have been averaged to hourly values.

Since planetary waves have periods of days, daily mean values for the $f_0F2$ frequency were produced that removed variations of the F-layer that occur with periods less than or equal to one day. These variations are mainly related to tides, photo-chemistry, ionization, and the rotation of the station under the auroral oval. To form daily mean values that were not influenced by these shorter period oscillations, the data analysis technique described in Kleinknecht et al. (2014) was employed.

This data processing takes place in two phases. In the first, the time series of data from an individual station is treated to obtain a daily station mean that is averaged over four days. In the second phase of data reduction, we examine these independent daily means from all the individual stations for a given day as a function of longitude, fitting spatial harmonics to the changes in longitude. To be identified as a planetary wave, the means of the individual stations must change in a periodic fashion with longitude, so that the variation of their means traces either a constant value ($S_0$), or one ($S_1$) or two ($S_2$) wavelengths around the globe.

In the first phase of data analysis, the hourly mean values at each station were divided into 4-day windows to provide a mean value averaged over 4 days. However, a simple boxcar or running-mean average would be skewed by the strong daily variation of ionization and the possible occurrence of a strong 2-day wave lying at the Nyquist frequency of our daily means. Thus, we fit the temporal harmonic components at 6 h, 8 h, 12 h, 24 h, and 48 h explicitly along with the mean value that represents the boxcar or running mean average of the residuals after the removal of the harmonic components. The window was then stepped in 1 day intervals and the process repeated to build up a time series of daily values for each station. To insure sufficient data coverage for the fit, 4-day windows that do not pass the quality control explained in Kleinknecht et al. (2014) were dismissed. Since the analysis for planetary waves uses daily means, the quasi-two-day wave that has been observed by e.g. Forbes and Zhang (1997), Pancheva et al. (1994) would lie at the Nyquist frequency. To prevent Doppler shifting by the mean winds aliasing this component, a length of 4-days was chosen for the windows to ensure adequate coverage of all daily periods 24 h or less, and to remove the quasi-two-day wave oscillation. Variations with periods less than four days will be effectively removed by the low pass filter imposed by the running-mean smoothing of the daily means over the 4-day window (Kennedy, 1980; Owens, 1978). Fig. 1 shows an example of the component fit to a 4-day window in January (05–08 January, 2001) from the ionosonde at King Salmon (KS759), while Fig. 2 shows the daily mean values for 2001 at all stations that are included in the study.

**Table 1**
Ionosondes between 52° and 65°N that produced suffcient available data during 2001 to be included in the study.

| Station name | latitude | Longitude E+.W- |
| --- | --- | --- |
| King Salmon (KS759) | 58.4 | −156.4 |
| College (CO764) | 64.9 | −147.8 |
| Gakona (GA762) | 62.4 | −145.0 |
| Goosebay (GSJ53) | 53.3 | - 60.4 |
| Narssarssuaq (NQJ61) | 61.2 | - 45.4 |
| Chilton (RL052) | 51.6 | - 1.3 |
| Juliusruh/Rugen (JR055) | 54.6 | 13.4 |
| Leningrad (LD160) | 60.0 | 30.7 |
| Moscow (MO155) | 55.5 | 37.3 |
| Novosibirsk (NS355) | 54.6 | 83.2 |
| Podkamennaya (TZ362) | 61.6 | 90.0 |
| Magadan (MG560) | 60.0 | 151.0 |
| Petropavlovsk (PK553) | 53.0 | 158.7 |

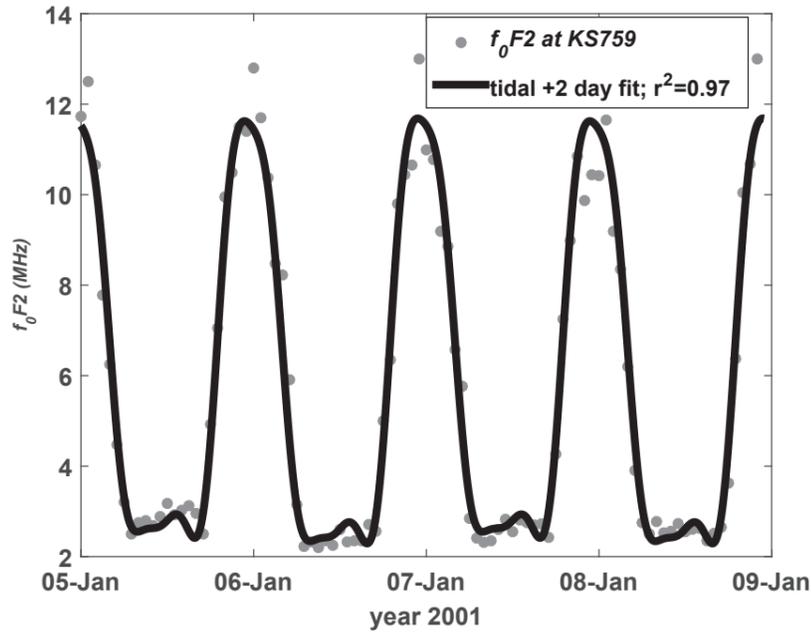

**Fig. 1.** Example of *tidal* fit (6 h, 8 h, 12 h, 24 h, and 48 h sine waves) to a 4-day window between 5th and 8th of January 2001. The hourly average raw data is shown as grey dots and the fit as a black solid line. The station abbreviations are listed in Table 1.

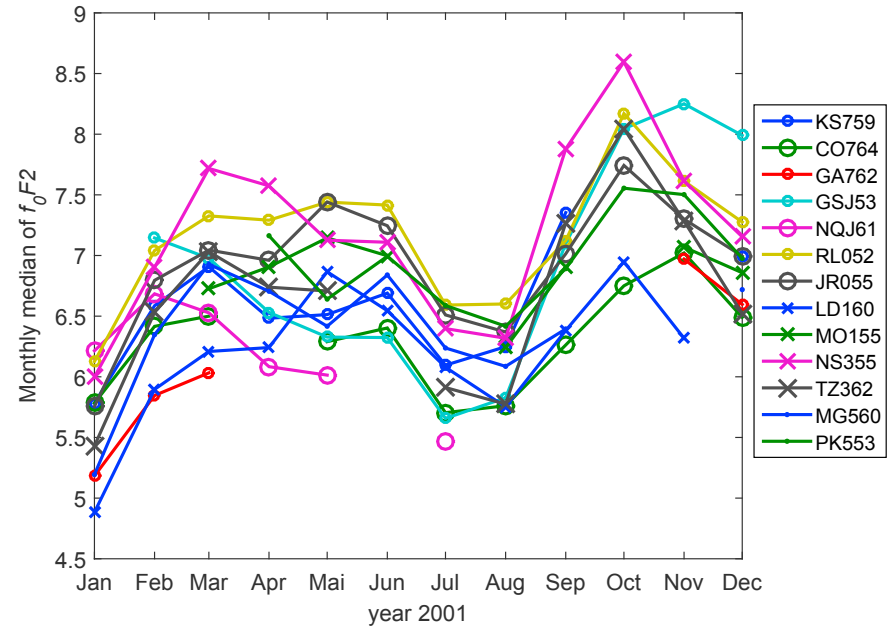

**Fig. 3.** Monthly mean $f_0F2$ values for 2001 at all stations included in the study. The station abbreviations are listed in Table 1.

In principle all ionosondes function in the same way and the data should be directly comparable. However, to remove possible bias between different stations due to the slight differences in latitude or operational offsets, the monthly medians have been removed for each station and each month (Araujo-Pradere et al., 2005). Fig. 3 shows the monthly median for each station. The relative variation between the stations' daily mean differences from their monthly medians, the $f_0F2$ anomaly, can then be used to derive the longitudinal structure of planetary wave-like oscillations.

For the second phase of data reduction, the extraction of the amplitude and phase of each longitudinal wavenumber, spatial harmonics were fit to the changes in longitude of the independent daily means from the individual stations, processed as described above, for each day for which there was sufficient data coverage. To determine whether a fit could be performed, the longitudinal chain was divided into six, 60° segments, and a fit performed if at least one of the stations in each segment had data. The fit of the daily anomalies of $f_0F2$ to the stations longitude extracted wavenumber 0, 1 and 2 and the phase of each wavenumber using the equation:

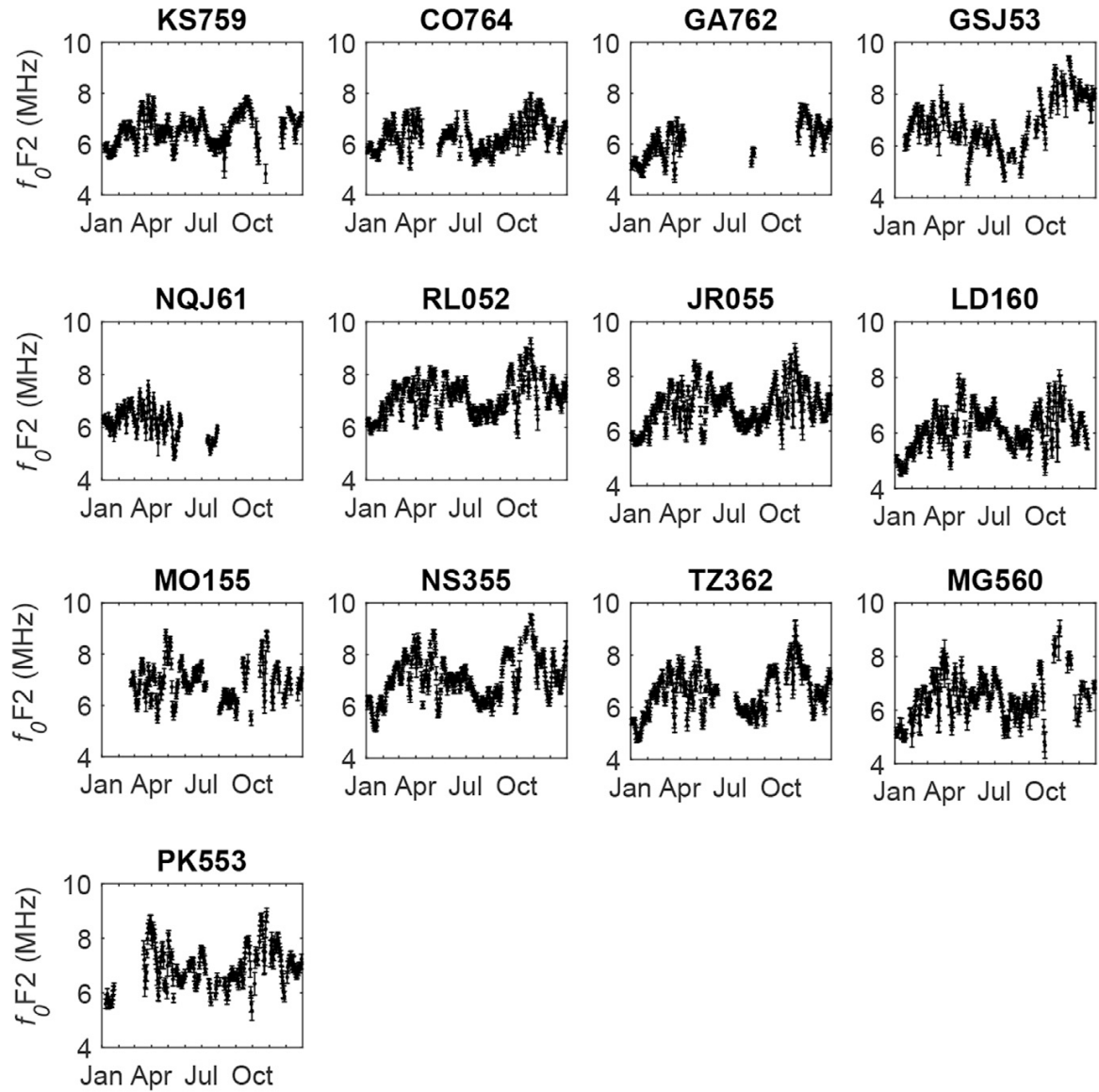

**Fig. 2.** Daily $f_0F2$ values for 2001 at all ionosonde stations included in the study. The station abbreviations are listed in Table 1.

$$SA_0 + SA_1 \cdot \sin\left(2\pi \frac{lon}{360} + c_1\right) + SA_2 \cdot \sin\left(2\pi \frac{lon}{180} + c_2\right) \quad (1)$$

Here $lon$ is the longitude in degrees, $SA_i$ is the amplitude of wavenumber $i$ and $c_i$ is the phase of the wavenumber $i$ at a given day (Kleinknecht et al., 2014). Fig. 4 shows an example of a longitudinal fit of the $f_0F2$ daily anomaly values for 21 November 2001. The blue dots show the value of the daily anomalies at the stations. The red line is the fit using equation (1) while the black, magenta and green lines show the amplitude and phase of wavenumber components 0, 1 and 2, respectively. As described above, any changes with periods of four days or less are filtered out of each station's data and will not cause relative offsets between the stations. While longer-period oscillations in solar flux change a station's mean value, all the stations are affected together so that the perturbations don't trace a periodic variation in longitude. Thus, they would not be interpreted as $S_1$ or $S_2$ oscillations. Hence, the analysis method used here is less susceptible to oscillations caused by solar flux or magnetic activity being interpreted as planetary waves. Performing the longitudinal fit at consecutive days makes it possible to observe the development of the different wavenumbers over time and hence gives the phase velocity of the planetary wave-like disturbances of each wavenumber.

Although the stations have a 13° spread in latitude (52°–65° N), the Hough functions characterizing the latitudinal structure of the individual temporal modes comprising each zonal wavenumber vary slowly. Using the same technique as Kleinknecht et al. (2014), this was tested by fitting planetary waves to wind re-analysis data at a single, average latitude and station longitude versus at the actual station locations. The resulting fitted amplitudes and phases agreed with each other to better than 10% and within the 2-$\sigma$ fitting uncertainties.

## 3. Results and discussion

A Hovmöller diagram is used to visualize the amplitude and phase propagation of planetary wave-like disturbances. In a Hovmöller diagram the fitted sinusoidal component for each wavenumber, $SA_i \cdot \sin\left(2\pi \cdot \frac{lon}{360} + c_i\right)$ is plotted over longitude along the horizontal, and each day's independent fit is stacked vertically. The amplitude variation in time and longitude is shown by colours, where positive and negative deviations are red and blue, respectively, with the saturation of the color indicating the magnitude of the fitted sinusoid. The Hovmöller diagram for the wavenumbers 0, 1 and 2 fitted to the daily anomaly from the monthly median $f_0F2$ values for the year 2001 is shown in Fig. 5. To account for the phase wrapping at $\pm 2\pi$ and better visualize the phase progression, the wavenumber one, $S_1$, results have been plotted over two longitudinal wave periods. The year 2001 was used for the analysis because it had enough stations for sufficient longitudinal fits to give a clear picture of the phase and amplitude behaviour throughout a whole year. This year also coincides with a year for which planetary waves in the MLT have been extracted for comparison (Kleinknecht et al., 2014).

To better examine the amplitude behaviour, Fig. 6 shows the amplitude of the ionospheric $S_0$, $S_1$, and $S_2$ components as a time series. There are occasions when all the wavenumber modes show similar behaviour, as has been observed in the mesosphere, stratosphere and troposphere. However, the $S_0$, which represents a simultaneous increase or decrease in $f_0F2$ at all stations, shows a large variability that may be related to modifications of the ionospheric density brought about by changes in solar flux or magnetic disturbances, as discussed in the next section. In this single year of data, there is also a large scatter in both the $S_1$ and $S_2$ amplitudes, but a distinct semi-annual variation can be discerned in both. However, while the $S_1$ component has a less pronounced maximum in autumn, the $S_2$ has large and sustained maxima in both spring and autumn.

During winter, the amplitude of the $S_1$ shows a clear intensification during a major stratospheric warming (SSW) event that occurred in late January and early February. While there is a similar enhancement in the $S_2$ amplitudes during this time, it is of the same order as other rapid variations in the data. By contrast, the $S_0$ decreases during this period, indicative of a decrease in the ionospheric density occurring at all stations. Although this decrease is of the same order as other variations seen in the $S_0$, the general decrease is consistent with the observed high-latitude electron-density response to stratospheric warming events that has been attributed to intensification of atmospheric tides (e.g. Goncharenko and Zhang, 2008; Pedatella and Forbes, 2010).

### 3.1. Neutral dynamics or magnetic disturbance?

A question arises as to what extent the oscillations in the $f_0F2$ frequency, which is proportional to the electron density in the F-region, are influenced by variations in auroral particle precipitation and solar flux. Emery et al. (2011) have found solar and magnetic variations with periods from 5 to 27 days which may influence the amplitude variation of $f_0F2$ at individual stations. Indeed, Altadill and Apostolov (2003) have found that up to 30% of the variance in $f_0F2$ at individual stations can be driven by geomagnetic effects. Similarly, Ma et al. (2012) show the 27 day solar rotation period is evident in the $N_mF2$ values from individual stations driven by both solar EUV and geomagnetic activity, with the geomagnetic forcing dominant at higher latitudes. To address this, a lagged correlation has been performed between the amplitude of the wavenumbers and the $A_p$, $k_p$ and $-Dst$ indices. Wavenumber zero, $S_0$, shows a significant moderate negative correlation at a 2–3 day lag (index leading). The comparison and the correlation between $k_p$ and $S_0$ is shown in the upper and lower plot in Fig. 7, respectively. The correlations for the other indices show similar behaviour.

The observed interplay between magnetic activity and the amplitude of the $S_0$ component can be related to a decrease in the daily average electron density associated with the high-latitude storm response (Buonsanto, 1999; Danilov, 2013). However, for the other wavenumbers no significant correlation was found at any specific lag indicating that there is no strong interplay between magnetic activity and the observed planetary wave-like perturbations observed in wavenumber one and two.

Although changes in solar flux affect all the stations together and do not manifest as a longitudinal variation in the daily means of the stations, the F10.7 index has been correlated with each longitudinal wave number to check for solar influence. Only the $S_1$ component shows a weak correlation coefficient of 0.3 at a lag of four days, with F10.7 leading. However, the $S_1$ is known to have temporal components extending to periods of 20 days (Forbes and Leveroni, 1992) that are similar to periods observed in F10.7 (Emery et al., 2011). Thus, the weak correlation may result from coincident but independent changes in F10.7 and the $S_1$ amplitude. Nevertheless, the low correlation, just significant at the 2-$\sigma$ level, indicates that the solar flux is not the primary driver of the observed planetary wavelike oscillations in the ionosphere.

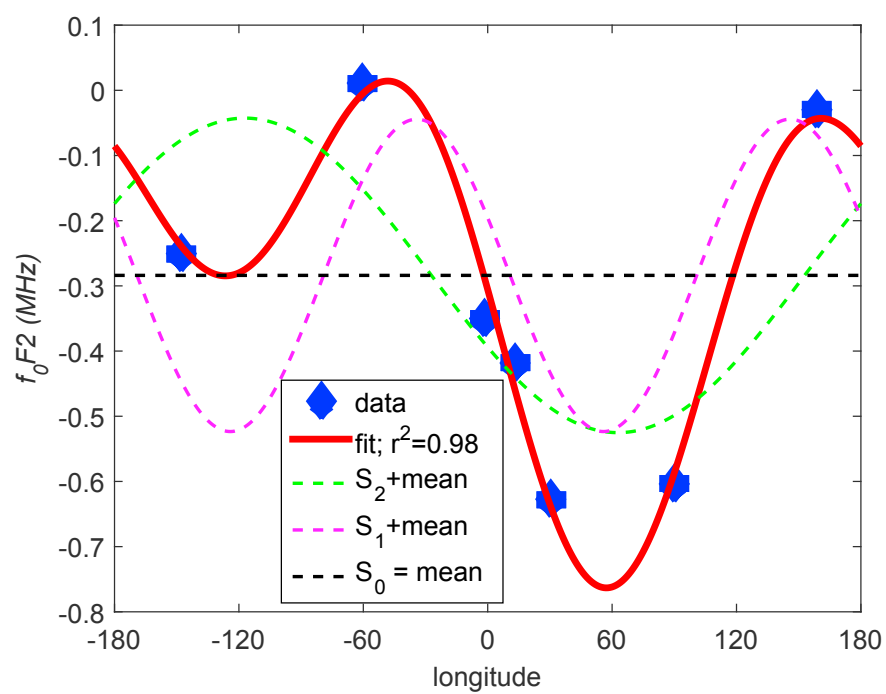

**Fig. 4.** Example of longitudinal fit. The daily anomalies at the different stations at the 21th October 2001 are shown in blue. The longitudinal fit is shown in red. Wavenumber zero, $S_0$, is black and wavenumbers one, $S_1$, and two, $S_2$, are shown off-set by $S_0$ in green and magenta, respectively.

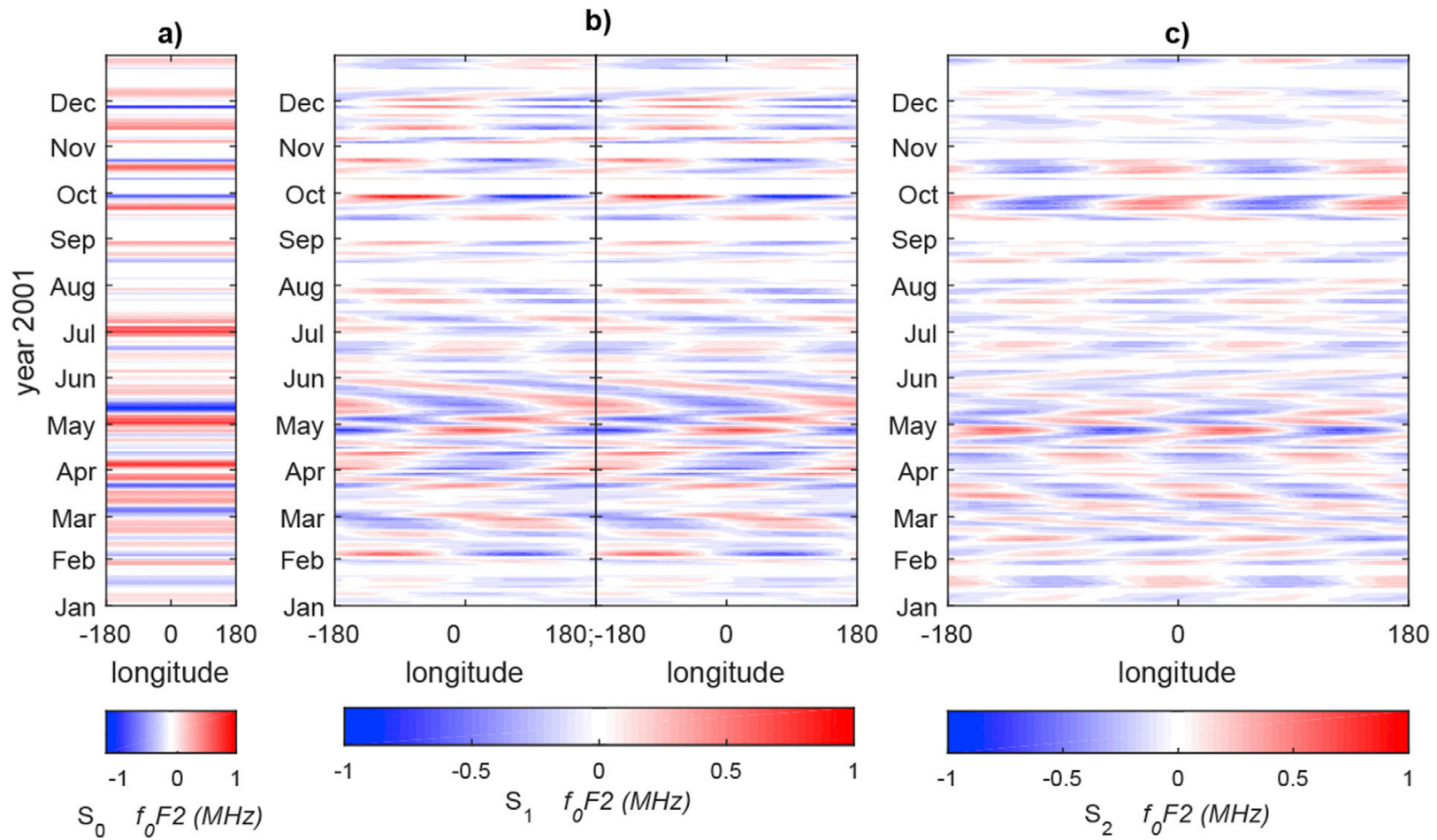

Fig. 5. Longitudinal wavenumbers $S_0$ (a), $S_1$ (b) and $S_2$ (c) of the daily monthly mean anomaly $f_0F2$ values during 2001. Red and blue colours signify positive and negative $f_0F2$ perturbations, respectively. $S_1$ is plotted over two longitudinal wave periods for better visualization.

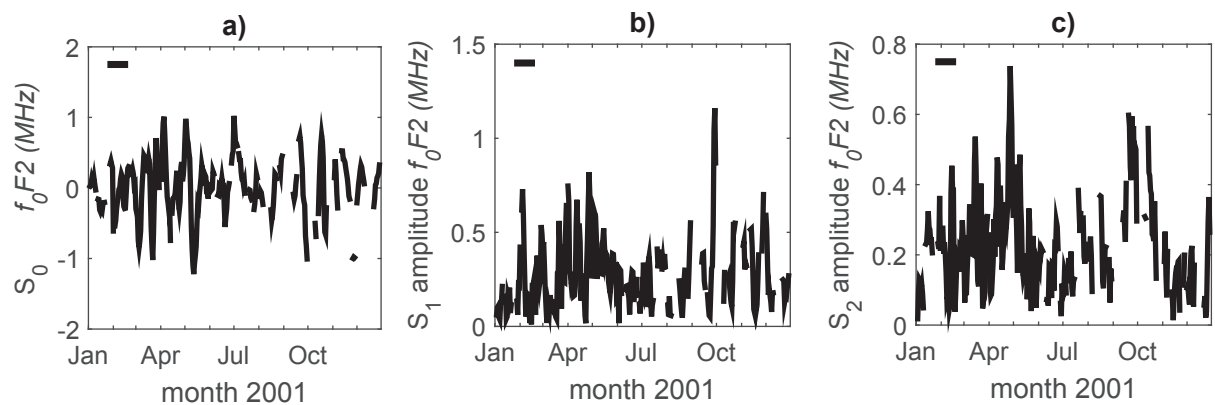

Fig. 6. Amplitudes of the longitudinal wavenumbers $S_0$ (a), $S_1$ (b) and $S_2$ (c) of the daily monthly mean anomaly $f_0F2$ values. The horizontal bar marks the sudden stratospheric warming that began on January 29 (Stray et al., 2015a).

The results presented here are at odds with work that has found substantial solar or magnetic influence on the temporal oscillations at planetary wave periods in the time series of hourly or noontime $f_0F2$ values at individual stations (e.g. Laštovička et al., 2003; Altadill and Apostolov, 2003). It could be that the total superposition of all temporal modes contained in the spatial modes observed with the longitudinal array of stations used here is less susceptible to these solar or magnetic perturbations than an individual temporal mode. However, it may be that the analysis used here is more resistant to periodic solar flux oscillations and recurrent magnetic storms that can imprint these periodicities, which are similar to planetary-wave periods (Emery et al., 2011), on the individual station's time-series of hourly or noontime $f_0F2$ values used in a Fourier analyses (Ma et al., 2012). For example, changes in the hourly or noontime $f_0F2$ values at a single station that were created by the 5-day solar-flux oscillation identified by Emery et al. (2011) could be interpreted as a 5-day planetary-wave oscillation whose amplitude was linked to that of the solar flux oscillation. However, in the spatial mode analysis used here, such solar or magnetic effects would have to cause perturbations to the daily mean $f_0F2$ values that have been averaged over four days, and these perturbations would need to trace out a periodic $S_1$ or $S_2$ variation in longitude across all the stations in order to be identified as a planetary wave. Thus, the spatial analysis used here may be more robust in discriminating against solar or magnetic perturbations to $f_0F2$.

Individual temporal modes have been observed to be highly variable in time (Day and Mitchell, 2010; Laštovička et al., 2003). This temporal variability is evident in the rapid amplitude and phase variations occasionally displayed in the $S_1$ or $S_2$ spatial modes observed in Fig. 5. Similarly, Altadill and Apostolov (2003) found that the temporal oscillations of $f_0F2$ did not appear at all the stations in a longitudinal chain. This would indicate that some of the stations used in the temporal analysis were located near the nodes of a spatial mode, where the superposition of all temporal modes in the $S_1$ or $S_2$ spatial modes interfere destructively and do not vary in time. Although the temporal periods at a given longitude can be recovered from the Fourier analysis of the time series of $S_1$ or $S_2$ amplitudes (Kleinknecht et al., 2014), this same ambiguity occurs for the spatial modes when the wave is stationary. Thus, the spatial and temporal analyses are complementary, and a combination of these techniques would be advantageous in future work.

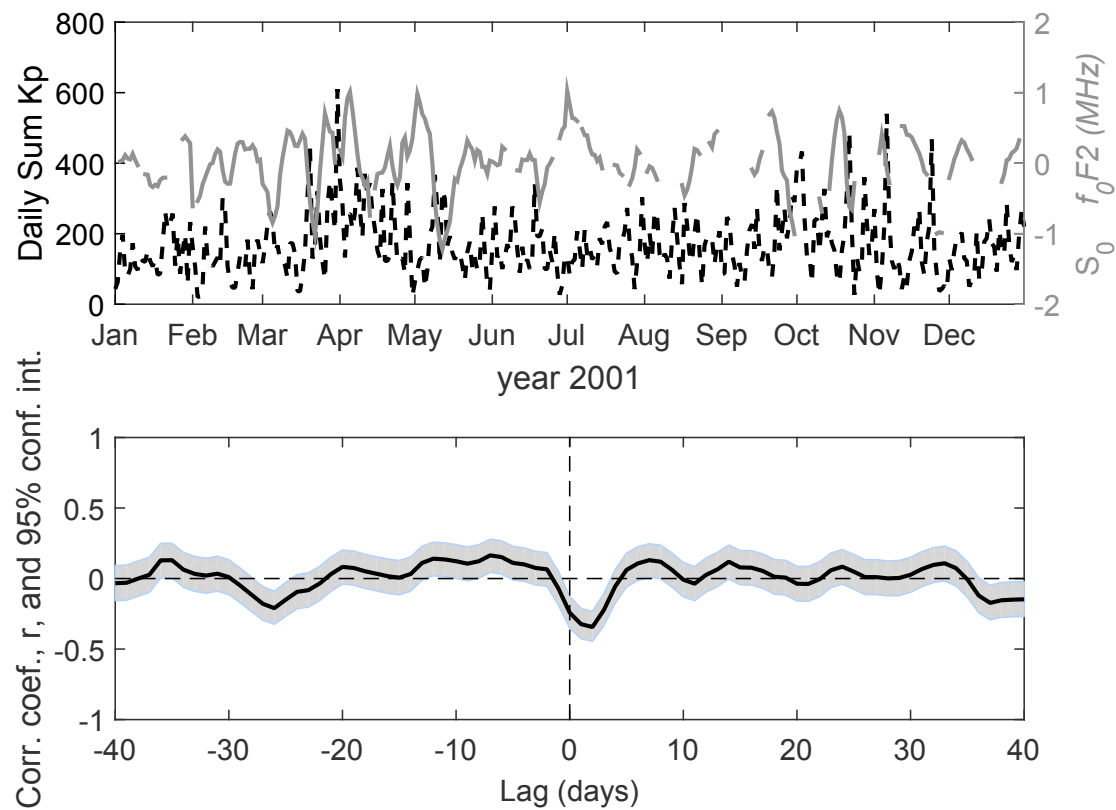

**Fig. 7.** (a) Comparison of magnetic index $k_p$ (dashed black) with wavenumber $S_0$ (solid grey) during 2001. Increasing phase indicates westward wave motion. (b) Lagged correlation of the $k_p$ and $S_0$ values shown in (a). The grey area marks the 95% confidence interval. Positive lags indicate that $k_p$ is leading.

### 3.2. Comparison with planetary wave activity in the mesosphere-lower thermosphere

Earlier studies (Kleinknecht et al., 2014; Stray et al., 2014; 2015a, 2015b) used a chain of SuperDARN radars to extract planetary waves of longitudinal wavenumber 1 and 2 from meridional wind in the mesosphere-lower thermosphere (MLT). The amplitudes of the $S_1$ and $S_2$ planetary waves in the meridional winds of the MLT during 2001 are taken from Kleinknecht et al. (2014) and shown in Fig. 8. The wave amplitudes display a high degree of scatter. Nevertheless, the weak annual cycle, peaking in winter, as well as the increases during spring and autumn that are apparent in the multi-year climatology of planetary waves given by Stray et al. (2014), can be discerned in the 2001 data. During 2001, the springtime peak is weak, but the September peak is strong.

To compare with the ionospheric oscillations shown in Fig. 6, we note that the amplitudes represent two different physical parameters, mesospheric wind in the MLT and $f_0F2$ in the ionosphere, which have no direct relationship. Thus, it is not possible to comment on the relative magnitudes of the oscillations and confine our remarks to changes in the amplitudes and phases (the longitudinal position of a wave maximum) throughout the year. It is clear that the weak annual cycle present in the MLT waves is not apparent in oscillations in $f_0F2$. While the intensifications near equinox occur in both regions, those in the ionosphere appear later (April and October) than those in the MLT (March and September).

One feature that appears coincident in the mesosphere and ionosphere is the change in wave amplitude that occurs after the onset of the stratospheric warming event in early February 2001 (Stray et al., 2015b). Thus, while the planetary wave response to the SSW is nearly simultaneous in the mesosphere and ionosphere, the seasonal effects in the ionosphere appear to be delayed relative to those in the MLT.

For better visualization of the propagating waves, a Hovmöller diagram of the MLT wavenumber 1 and 2 waves is shown in Fig. 9 together with those from the ionosphere, again with wavenumber one plotted over two longitudinal wave periods for clarity. While the magnitudes of their amplitudes differ, as discussed above, it is apparent that the large scale behaviour of their phase motions (their eastward or westward movement) appears to be linked together.

For the $S_1$ components, there appears to be a temporal offset of the phase motion, similar to that seen in the amplitudes. To demonstrate, the fitted phases of the $S_1$ components in the MLT and the ionosphere, giving the longitudinal positions of the wave structures, are shown in Fig. 10a. Both components undergo a westward and eastward oscillation at the start of the year. Then both begin a slow, sustained, westward motion that is followed by a sustained period where both waves become nearly stationary. However, it is clear that the wave motion in the MLT leads that in the ionosphere by about 30 days, as was seen in the seasonal variations of the amplitudes. To quantify this, a lagged correlation was performed and is presented in Fig. 10b. Here we see a high, significant correlation of the waves in the MLT and the ionosphere, with a correlation coefficient of $0.90 \pm 0.02$ at a 30-day lag, indicating that the MLT wave motions lead those in the ionosphere by 30 days.

The observed 30-day delay between wave motions and amplitude near 100 km (MLT) and 300 km (ionosphere) represents a vertical propagation speed of about 6–8 km/day (5–8 cm/s). This would indicate that the mechanism transferring the MLT wave signatures into variations of the ionospheric $f_0F2$ value is indirect. Pancheva and Lysenko (1988) suggested that vertical motions associated with planetary wave oscillations near the turbopause ($\approx$100–105 km) could induce changes in the $O/N_2$ and $O/O_2$ ratios, affecting recombination rates and, hence, F-region electron densities. Assuming planetary waves cease propagating near the turbopause due to diffusion, any perturbations in the $O/N_2$ ratio caused by MLT planetary waves would propagate only at the diffusion speed of O-density perturbations into the ionosphere. Due to slow diffusion velocities, on the order of 0.1–5 cm/s (≈0.1–5 km/day) below 120 km (Johnson and Gottlieb, 1973), the observed 30-day lag is commensurate with the diffusion time for a planetary-wave composition changes to propagate from the MLT near the turbopause to reach the ionosphere.

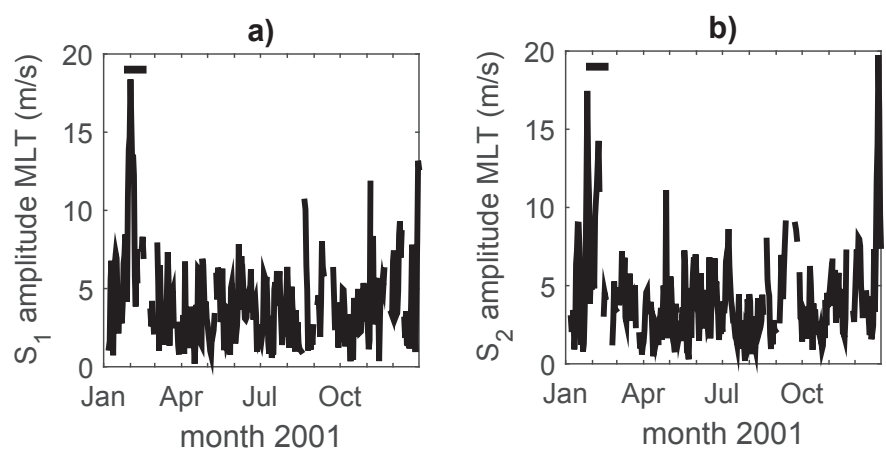

**Fig. 8.** Amplitudes of the longitudinal wavenumbers $S_1$ (a) and $S_2$ (b) of the daily anomaly neutral mesospheric wind observed by a chain of SuperDARN radars in the mesosphere lower-thermosphere (MLT) (Kleinknecht et al., 2014). The horizontal bar marks the sudden stratospheric warming that began on January 29 (Stray et al., 2015a,b).

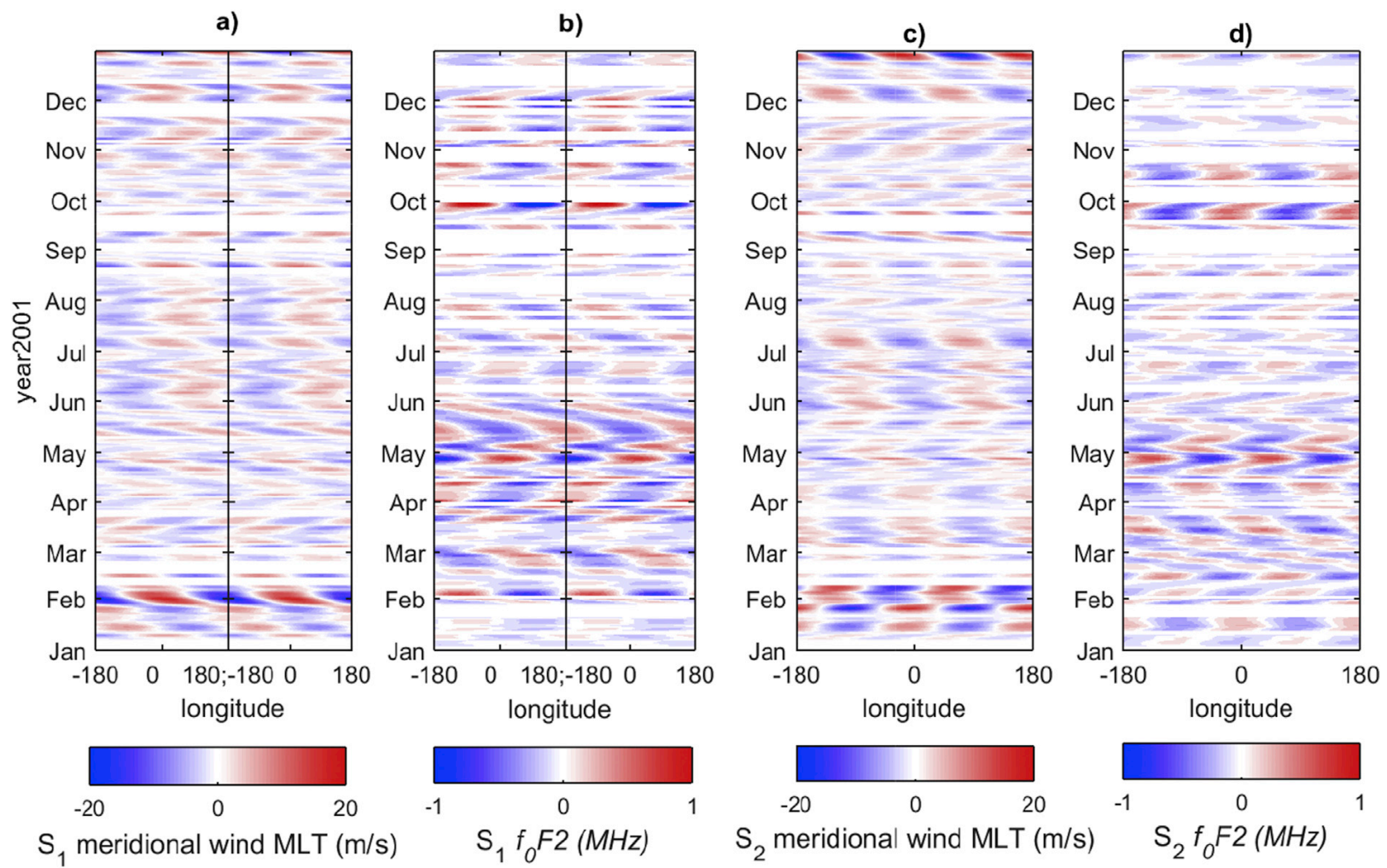

Fig. 9. Longitudinal wavenumbers $S_1$ and $S_2$ in the MLT (a, c) and in the ionosphere (b, d). Red and blue colours signify positive and negative perturbations, respectively.

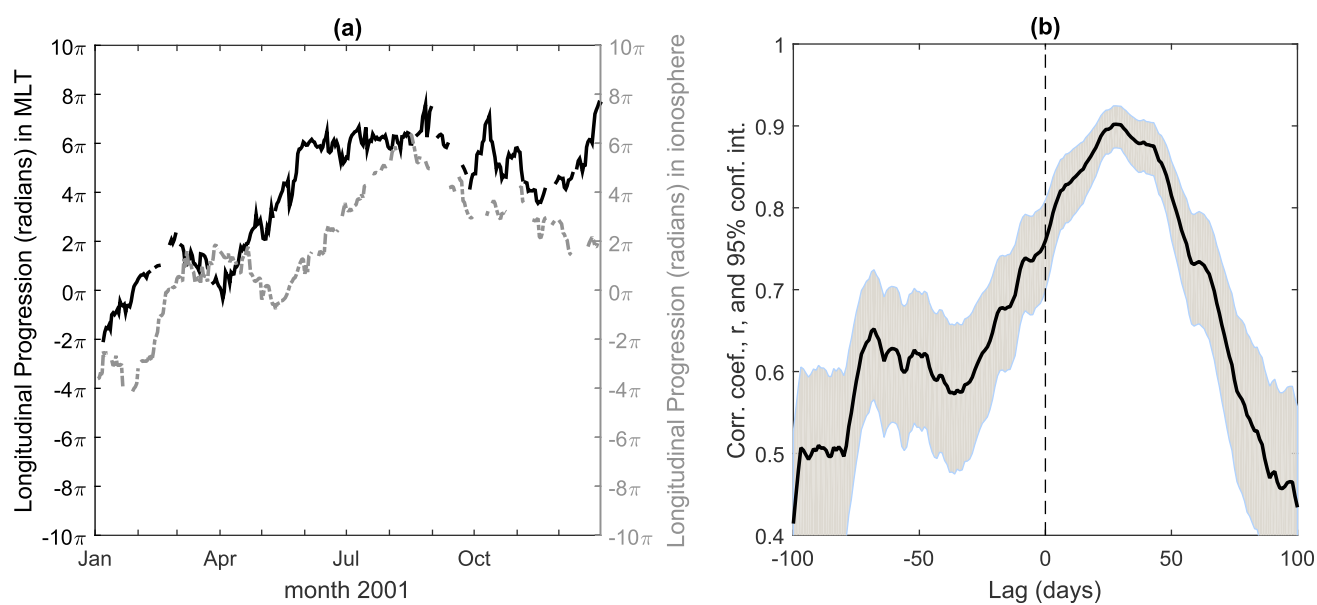

Fig. 10. (a) Comparison of the longitudinal phase progression of wavenumber one, $S_1$, observed in the MLT (solid black) by a chain of SuperDARN radars and in the ionosphere (dashed grey) observed by a chain of ionosondes. (b) Lagged correlation of the longitudinal phase progression in the MLT and ionosphere shown in (a). The grey area shows the 95% confidence interval. Positive lags indicate that the MLT is leading.

These results show presence of an $S_1$ structure in the ionosphere that is only weakly influenced by solar or particle precipitation effects. This structure is observed to move synchronously with the $S_1$ planetary wave in the MLT at a fixed, 30-day time delay, which is commensurate with the diffusion time of O density perturbations to reach the ionosphere from the turbopause. Taken together, this supports the suggestion that the global oscillation in $f_0F2$ is coupled to the neutral dynamics below through the diffusion mechanism suggested by Pancheva and Lysenko (1988).

The fitted phases of the $S_2$ components in the MLT and the ionosphere, giving their longitudinal position, are shown in Fig. 11a. In comparison with the $S_1$, both components are relatively stable but undergo two periods of motion during spring and autumn, with the motion persisting and more pronounced in the ionosphere. The largest motions occur in the autumn, where the motion in the MLT occurs rather early and peaks at equinox. However, the motion in the ionosphere only starts at equinox and peaks much later. By the time the ionospheric structure has reached it maximum westward position, the MLT structure has moved to its farthest eastward position.

The lagged correlation is performed and are presented in Fig. 11b. The cross-correlation function shows a sinusoidal variation with a period of approximately 180 days, or a semi-annual variation, and indicates that the variations in the MLT and the ionosphere are out of phase. In the MLT, the seasonal change of stratospheric wind direction, caused by changes in solar heating of ozone, have been shown to initiate this rapid motion of the planetary wave structure, particularly in the autumn, and can give rise to a semi-annual variation (Liu et al., 2001; Stray et al., 2014). The high correlation between motion of the $S_2$ structures in the MLT and ionosphere near ± 90 days (≈0.7±0.1) suggests that the ionospheric waves also undergo a the semi-annual variation, possibly associated with the semi-annual variation of F-region winds near 300 km (Emmert et al., 2003, 2006; Forbes, 2007; Fuller-Rowell, 1998). However, one year of data is insufficient to investigate this suggestion further.

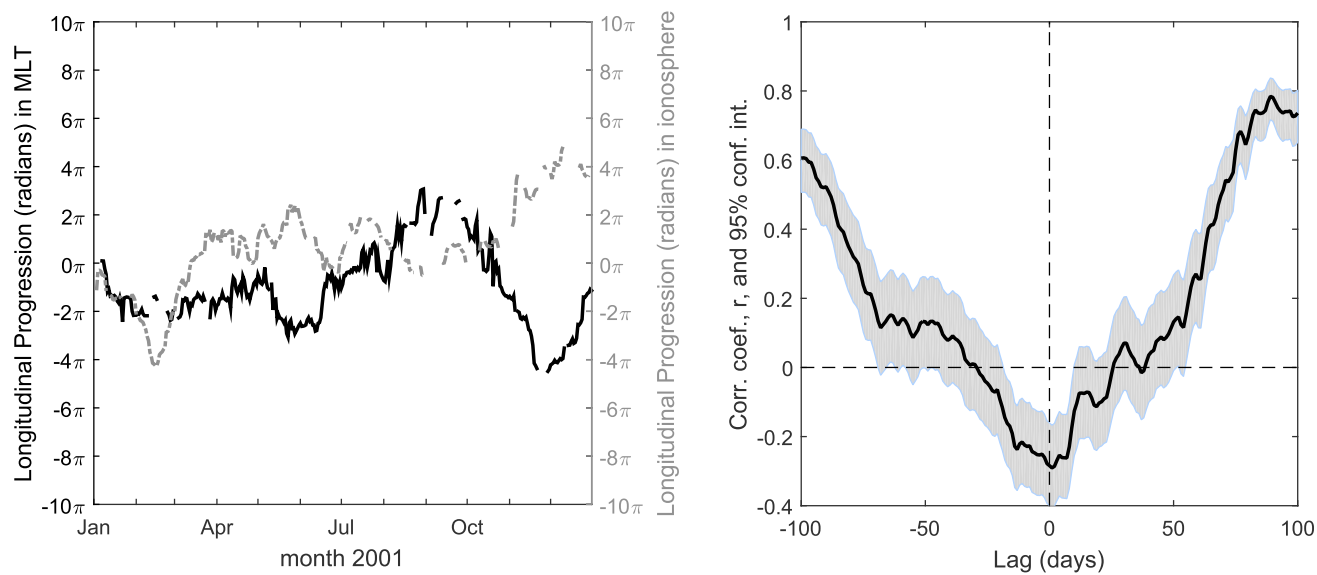

**Fig. 11.** Like Fig. 10, but for wavenumber two, $S_2$.

### 4. Conclusions

A longitudinal chain of ionosondes in a latitude band between 52° and 65° N has been used to examine planetary wave-like oscillations in $f_0F2$, which is proportional to electron density in the F-region, and their possible coupling to the planetary wave activity in the MLT during 2001 as measured by Kleinknecht et al. (2014). In contrast to other studies that examine individual velocity modes of a single wavenumber component (i.e. different temporal periods), the method employed here, an adaptation of Kleinknecht et al. (2014), captures the total wavenumber component inclusive of all temporal periods. Examining the total wavenumber structures gives insight into possible mechanisms for communicating the planetary wave information from the lower atmosphere into the ionosphere since many depend on the net wind or density structure formed by the superposition of all the temporal modes.

The geographic extent of the available ionosonde chain is sufficient to retrieve planetary scale structures in the ionosphere with wavenumbers zero ($S_0$), one ($S_1$), and two, ($S_2$). Comparison and correlation of the observed planetary wave-like oscillations to indices of magnetic activity indicate that only the $S_0$, representing a zonal decrease in the daily mean electron density, is driven by magnetic activity. In addition, the ionospheric oscillations could be compared to $S_1$ and $S_2$ planetary wave structures in the MLT. The wave amplitudes in both regions increase simultaneously in response to a stratospheric warming event. However, while both show a similar behaviour near equinox, the changes in $f_0F2$ are stronger and occur later than those in the MLT. In addition, the phase variations of the $S_1$ components appear to move synchronously, with the waves in the ionosphere again lagging the wave motions in the MLT. To quantify this, the cross correlation of the phase propagation between the MLT and the ionosphere show a strong correlation (0.9±0.2) at a lag of 30 days, with the MLT motions leading those in the ionosphere. This delay is commensurate with the propagation speed at which a change in the $O/N_2$ density would diffuse from the MLT to the ionosphere. On the other hand, the phase variations of the $S_2$ components in the MLT and ionosphere anti-correlate at zero lag and correlate at ±90 days, indicating a semi-annual oscillation that may be related to semi-annual variations in the mean wind in the MLT and ionosphere.

The results presented here show presence of an $S_1$ and $S_2$ structures in the ionosphere whose amplitude variations lag those observed in the MLT and which are only weakly influenced by solar or particle precipitation effects. The 30-day time lag between motions of the $S_1$ wave in the MLT and the ionosphere is commensurate with the time taken for O density perturbations from planetary waves near the turbopause to diffuse to the ionosphere, in agreement with the diffusion mechanism suggested by Pancheva and Lysenko (1988). In addition, the $S_2$ ionospheric structures undergo rapid migration as the winds change near equinox at the same time as those in the MLT, albeit in the opposite direction, in agreement with observations and models(Stray et al., 2014; Liu et al., 2001). Taken together, these results indicate that the wave structures observed in the ionosphere are connected with those observed in the MLT. While the $S_1$ structures appear to be related to the upward diffusion of MLT density perturbations, the mechanism coupling the $S_2$ structures, while indicative of a more direct process, remains uncertain.

To better understand the exact mechanisms by which the $S_1$ planetary wave in the MLT transfers information to the ionosphere, more years of ionosonde data with sufficient temporal and spatial data coverage would be needed. Recent work using $f_0F2$ derived from total electron content measured from satellites may benefit from treating the total wavenumber components formed by the superposition of the individual temporal periods. Finally, it is important that these ionospheric components be compared with robust measurements of planetary wave activity in the lower atmosphere that are not compromised by spatial-temporal aliasing.

### Acknowledgments

This study was supported by the Research Council of Norway/CoE under Contract 223252/F50. The authors acknowledge the use of ionospheric data from the Space Physics Interactive Data Resource, and wish to thank J. Gjerloev and R. Barnes for their helpful discussions.